\documentclass{aa}
\usepackage{psfig}

\def\G{$\Gamma_{\rm x}$ }

\def\approxlt{\mathrel{\hbox{\rlap{\lower.55ex \hbox {$\sim$}}
        \kern-.3em \raise.4ex \hbox{$<$}}}}
\def\approxgt{\mathrel{\hbox{\rlap{\lower.55ex \hbox {$\sim$}}
        \kern-.3em \raise.4ex \hbox{$>$}}}}

\begin{document}

 \title{Supersolar metal abundances and the Broad Line Region  
        of Narrow-line Seyfert\,1 galaxies} 


 \author{Stefanie Komossa\inst{1}, Smita Mathur\inst{2}}     

 \offprints{St. Komossa, \\
  skomossa@xray.mpe.mpg.de}

 \institute{
    Max-Planck-Institut f\"ur extraterrestrische Physik,
         Giessenbachstr., D-85748 Garching, Germany 
 \and 
 The Ohio State University, 140 West 18th Avenue, Columbus, OH 43210, USA }

 \date{Received: February 2001; accepted: June 2001}

   \abstract{Narrow-line Seyfert\,1 galaxies (NLSy1) are intriguing
due to their extreme continuum and emission line properties
which are not yet well understood.  This paper is motivated by
the recent suggestion that NLSy1 galaxies might be young
active galactic nuclei (AGNs) and that they may have
supersolar metal abundances. We have examined the stability of
broad emission line region (BLR) clouds under isobaric
perturbation and its dependence on gas metal abundances. We
show that supersolar metallicity increases the range where
multiple phases in pressure balance may exist. This is
particularly important for NLSy1s with steep X-ray spectra
where it delays the trend of hindrance of a multi-phase
equilibrium.  Finally, the role of warm absorbers in the
context of this scenario is assessed.   } 


\authorrunning{St. Komossa, S. Mathur}
\titlerunning{The BLR of NLSy1 galaxies}

   \maketitle

\vspace*{-11.2cm}
\Large
\begin{verbatim}
    RESEARCH NOTE
    \end{verbatim}
\vspace*{8.5cm} 
\normalsize
%
\section{Introduction}

\subsection {Narrow-line Seyfert\,1 galaxies}

X-ray and optical observations of the last decade revealed a new
sub-class of active galaxies, termed Narrow-line Seyfert\,1 galaxies
(NLSy1s hereafter).  NLSy1 galaxies are intriguing due to their
extreme continuum and emission line properties. Their optical broad
lines are narrower than in `normal' Seyferts and they show strong FeII
emission. Their X-ray spectra are often unusually soft.  Many
properties of AGN were shown to correlate strongly with each
other. The strongest variance, often referred to as `eigenvector 1' is
defined by the correlation between the width of the H$\beta$ emission
line and the strength of the [OIII] emission line, and the
anticorrelation with the ratio FeII/H$\beta$ (Boroson \& Green 1992).
NLSy1 galaxies are placed at one extreme end of eigenvector 1 and, in
general, their X-ray softness correlates with their optical
properties.  There is evidence that the X-ray properties of NLSy1
galaxies are best explained by a high accretion rate close to the
Eddington rate (Pounds et al. 1995).  Recently, it has been proposed
that the reason for the high accretion rate is the age (Mathur 2000a).
In this scenario NLSy1 galaxies with high accretion rate are AGNs in
an early evolutionary phase.  There are several lines of evidence
which indicate that NLSy1 galaxies may have super-solar gas phase
metal abundances (Mathur 2000a,b; see our Sect. 3 for details) and
thus represent the analogue to high-redshift quasars which are
believed to be young objects, and which show supersolar metallicity as
well (e.g., Hamann \& Ferland 1993, Dietrich et al. 2000).

In this {\sl Note}, we discuss the influence of
supersolar gas phase metal abundances on the possible
multi-phase equilibrium of the broad-line clouds in NLSy1 galaxies.

\subsection {Multi-phase cloud equilibrium: previous approaches} 

Different mechanisms for the confinement/stability of the broad line
clouds have been studied over the years.  The model originally
investigated by Krolik, McKee \& Tarter (1981; KMT hereafter) was a
two-phase BLR cloud model, consisting of cold line-emitting clouds ($T
\simeq 10^4$ K) in pressure balance with a hot inter-cloud medium ($T
\simeq 10^8$ K).  The original KMT model faces problems when a more
recent AGN continuum shape is used, in the sense that a pressure
balance between a cold, photoionization heated, and a hot, Compton
heated phase no longer exists.  In addition, depending on the
continuum shape, the hot phase may have lower temperature ($T \simeq
10^7$ K) and produce observable signatures due to its
 optical thickness 
(e.g., Elvis et al. 1985, Fabian et al 1986,
Rees 1987).  

However, more recent studies have established the presence of an
additional stable region of intermediate temperature (Reynolds \&
Fabian 1995, Komossa \& Fink 1997a,b), and 
the recent discovery of warm absorbers located in that
intermediate region have revived the interest in multi-phase cloud
models.  It is also interesting to note that recent {\sl Chandra} HETG 
observations of NGC\,4151 confirmed the existence (Elvis et al. 1990) of a high-temperature
X-ray emitting plasma, interpreted   
in terms of a
hot intercloud medium (Ogle et al. 2000).
This observation can be well explained by the 
classical KMT modeld, due to the very flat X-ray spectrum
of NGC\,4151 (Komossa 2001b). 

With respect to Narrow-line Seyfert\,1 galaxies, Brandt et al. (1994)
suggested that the relatively narrow optical BLR lines of these
galaxies might be traced back to the hindrance of BLR multi-phase
equilibrium in the presence of {\em steep} X-ray spectra. Model
calculations of Komossa \& Meerschweinchen (2000, KM2000 hereafter;
see also Komossa \& Fink 1997a) confirmed the idea. 
However, 
recent observations suggest that other mechanisms might be
at work (e.g., a small black hole mass would lead to narrow
lines in NLSy1 galaxies).  
Motivated by the recent evidence for supersolar metal
abundances in NLSy1 galaxies we extend here the earlier calculations to
explore in more detail the influence of gas abundances. We show that
supersolar metallicity increases the range where multiple phases in
pressure balance may exist, and that it delays the trend of hindrance
of a multi-phase equilibrium due to steep X-ray spectra.

\section{Model calculations and results} 

\subsection{The model}

We used the photoionization code {\em Cloudy} (Ferland 1993) to
carry out the calculations.  In order to obtain the cloud temperature
as a function of pressure the following assumptions about the cloud
properties were made:
 
Constant gas density ($\log n_{\rm H}=9.5$)
and solar element abundances according to Grevesse \& Anders (1989) were adopted,
and then varied relative to the solar value. 
It was assumed that there is no strong temperature gradient
across the clouds.
The clouds were
illuminated by the continuum of a central point-like energy source
with a spectral energy distribution (SED) from the radio to the
gamma-ray region as in Komossa \& Schulz (1997). This mean Seyfert
SED{\footnote{Rodriguez-Pascual et al. (1997) collected IR $-$ X-ray
fluxes of Seyfert and NLSy1 galaxies and concluded that both are
generally very similar concerning luminosities in different energy
bands except that NLSy1s tend to be underluminous in the UV.}}
consists of piecewise powerlaws with, in particular, an energy index
$\alpha$$_{\rm uv-x} = -1.4$ in the EUV.  The X-ray photon index
$\Gamma$$_{\rm x}$ was varied to cover the range typically observed in
NLSy1 galaxies.

The cloud temperature after each model run was extracted,
 as a function of the ionization parameter. 
The ionization parameter is defined as 
    \begin {equation}
     U = {Q \over {4\pi{r}^{2}n c}}~~, 
    \end {equation}
where $n$ is the gas density, $r$ the distance of
the gas cloud from the continuum source, $c$ the speed
of light, and $Q$ the number rate of photons
above the Lyman limit ($\nu_0 = 10^{15.512}$ Hz).

The stability of broad line clouds to isobaric perturbations can be
examined by studying the behavior of temperature $T$ as a function of
pressure.  If the temperature is multi-valued for constant pressure,
and the gradient of the equilibrium curve is positive, several phases
may exist in pressure balance.  Results are shown in Figure 1 where variations 
in $U/T$ correspond to variations in $1/{pressure}$.
 In the upper part of Fig.1, we have plotted  a sequence of
curves which were derived by using different X-ray slopes in the SEDs
which illuminate the clouds.  The powerlaw index changes from \G =
--1.5 to --4 from left to right, and the multi-valued nature of the
curves disappears for steep X-ray spectra.  Due to the relatively
weaker X-ray flux and the fact, that the value of $U$ is dominated by
the EUV flux near the Lyman limit, the gas remains longer in the phase
which is characterized by photoionization-heating and
collisional-excitation-cooling.  The same holds for a continuum with a
hot soft X-ray excess (KM2000). For comparison with results obtained
using the mean Seyfert SED, we have collected multi-wavelength
continuum observations of the NLSy1 galaxy NGC\,4051 (Komossa \& Fink
1997a and references therein), and plotted the corresponding
equilibrium curve in Fig. 1 (see next Section for details).

Metal abundances, which affect the cooling of the gas, were then
varied.  We find that supersolar metal abundances increase the range
where a multi-phase equilibrium is possible, whereas subsolar
abundances have the opposite effect (cf. the curve of Fig. 1 which
refers to \G =-- 1.9).  Whereas SEDs with an X-ray spectrum steeper
than \G $\approx -2.5$ do no longer allow for multiple phases in
pressure equilibrium if {\em solar} gas phase abundances are adopted,
such equilibria are still possible in case of supersolar abundances.

  \begin{figure*}
 \psfig{file=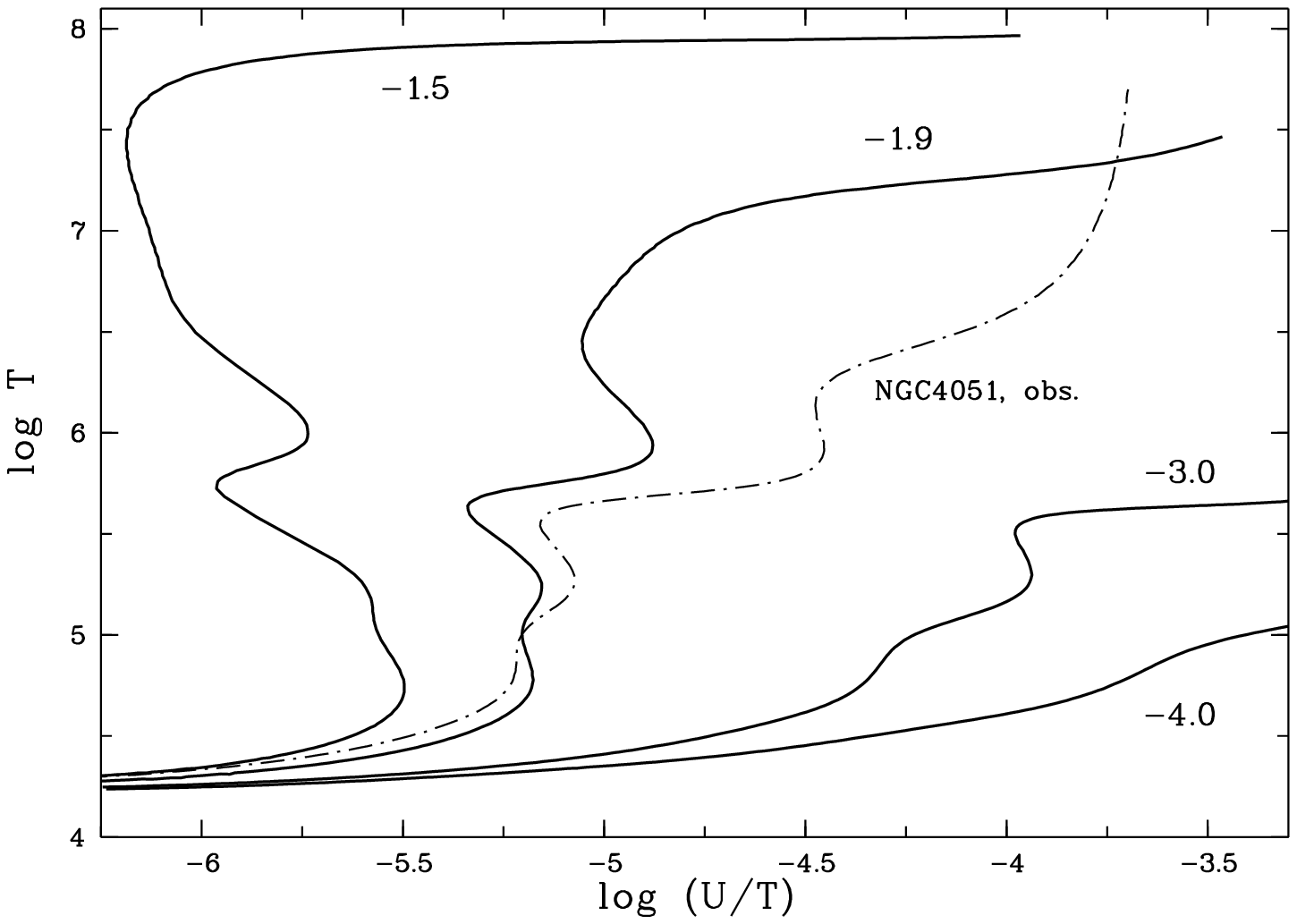,width=8.6cm,clip=}
 \psfig{file=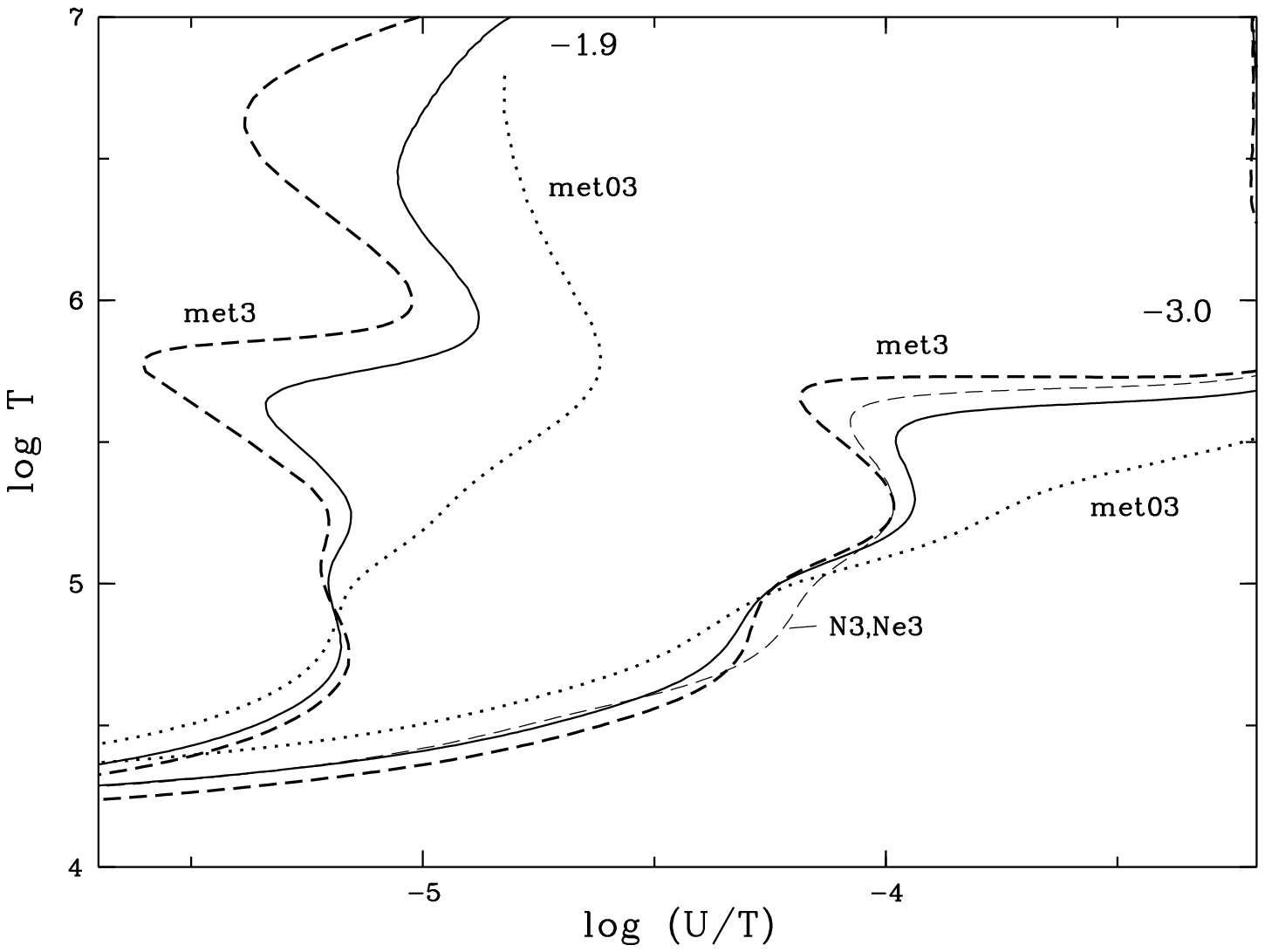,width=8.6cm,clip=}
 \psfig{file=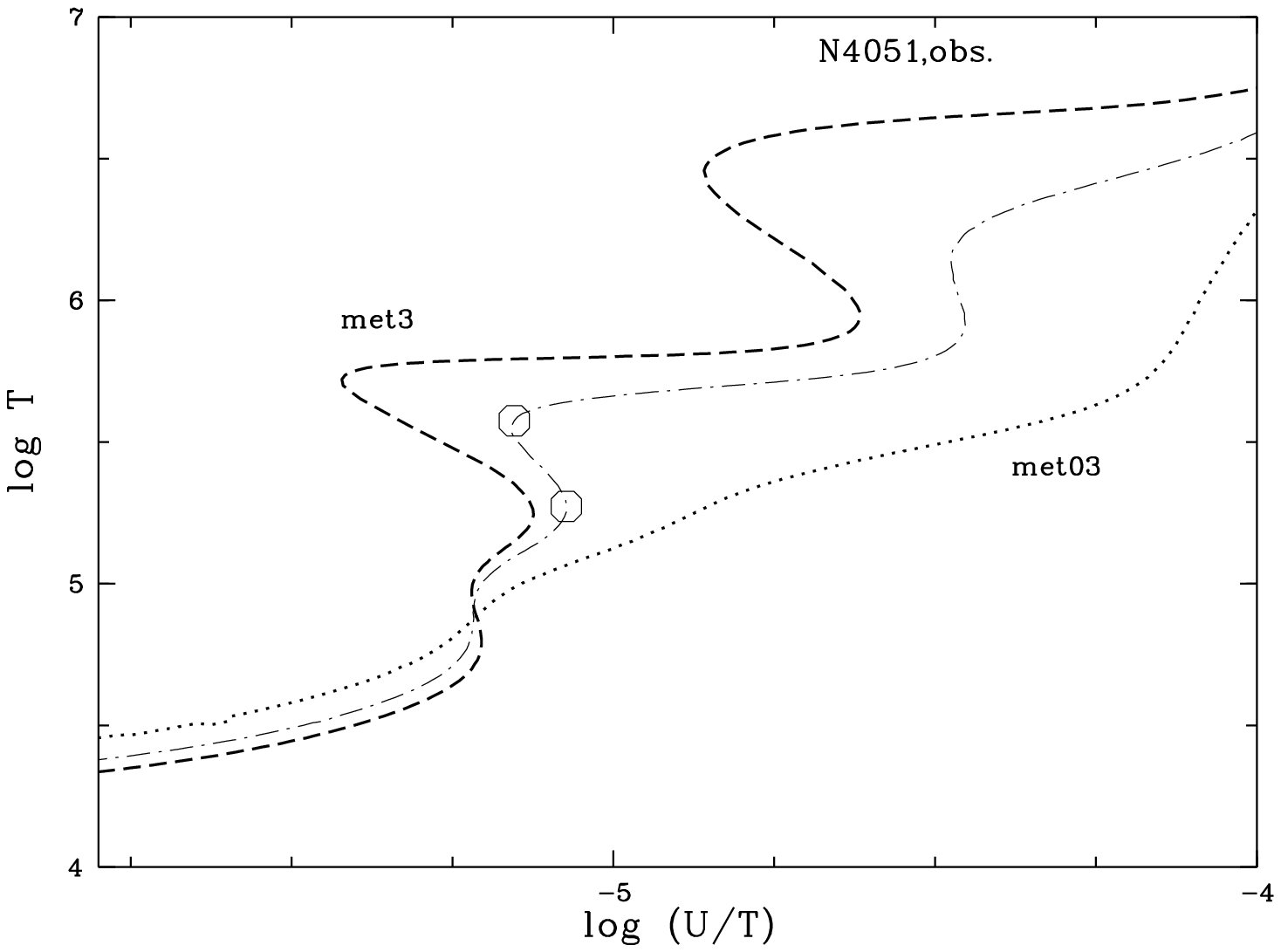,width=8.6cm,clip=}
 \caption[eq]{Equilibrium gas temperature $T$ against $U/T$
for various SEDs incident on the gas and metal abundances of the gas.
{\bf Upper panel}: 
Solid black lines: mean Seyfert input continuum with \G = --1.5, --1.9, --3.0 
and --4.0 as labeled in the graph; dot-dashed line: observed
multi-wavelength continuum of the NLSy1 galaxy NGC\,4051 with
\G = --2.3. Solar gas phase abundances were adopted.  
{\bf Middle panel}: sector of upper panel, showing the 
dependence on abundances. Solid lines: mean Seyfert input continuum with \G = 
--1.9
and --3.0 replotted for comparison; 
thick dashed lines: overabundant metals ($Z = 3 \times Z_{\odot}$),
thin dashed line: selectively enhanced nitrogen and neon abundances
of $3 \times$ solar, 
dotted lines = underabundant metals ($Z = 0.3 \times Z_{\odot}$).
{\bf Lower panel}:
results specific to NGC\,4051. The dot-dashed line corresponds, again,
to the observed multi-wavelength continuum of NGC\,4051 using 
solar abundances. 
  The open circles mark the location of NGC\,4051's 
   warm absorber, observed at two different
  epochs (Nov. 1993, Komossa \& Fink 1997a; and Nov. 1991, KM2000).
  The values were derived from one-component
  warm absorber fits to the {\sl ROSAT} X-ray spectrum of this galaxy,
  assuming solar metallicity.  
Dashed and dotted line have the same meaning as above. 
}
\label{eq}
\end{figure*}

In a next step, we selectively increased the neon and nitrogen
abundances ($Z_{\rm N,Ne} = 3 \times Z_{\odot}$), motivated by
(a$_{1}$) the observations of X-ray spectral features around $\sim$
1.1-1.4 keV which could partly be due to increased Ne absorption,
(a$_{2}$) optical evidence for overabundant N, and (b) the report of
Fabian et al. (1986) that Ne has a particularly strong influence on
the equilibrium curve (but see comment (ii) of our Sect. 3.3).  We do
not find any exceptional effects (Fig. 1), but note that modifications
of the shape of the curve are dominated by Ne, not N.

\subsection{The BLR in NGC 4051}

  We have constructed the equilibrium curves with the SED of the NLSy1
  NGC~4051 as an input. The results for solar and super-solar
  abundances are shown in Figure 1c. 
Below, we give some order-of-magnitude estimates of the parameters
of the BLR of NGC\,4051 in the context of our general modeling 
approach.    
  It is clear from the figure that a
  two-phase equilibrium with common pressure does not exist for solar
  abundances. On the other hand, such an equilibrium is possible if
  abundances are super-solar. In the context of the BLR cloud model,
  the lower temperature region corresponds to the BLR clouds while the the
  confining medium is in the hot region. From Figure 1c, we find that
  the ionization parameter of the BLR spans a range between $\log
  U=0.12$ and $\log U=0.30$. Reverberation mapping observations can
  determine the distance of the BLR from the nucleus accurately. In
  NGC~4051 it was found to be $1.6 \times 10^{16}$ cm (Peterson et al.  
  2000). Using equation 1, the density in the BLR can be constrained to
  be $8.7<\log n< 9.1$.{\footnote {This density
  estimate is based on a value of $Q = 1.6\,10^{52}$ 1/s of
  NGC\,4051, obtained using a piecewise powerlaw as SED. 
 `Photon counting arguments' based on the H$\beta$ luminosity
  suggest that this is a lower 
  limit on $Q$ (Komossa \& Fink 1997a).
  A higher value of $Q$ would increase
  the inferred density} }   
  If such a model for a BLR is valid for all NLSy1s, then it implies
  that BLR densities in NLSy1s are low and ionization parameters
  are higher than the ``standard'' value ($\log U \approx -2$). 
  Higher densities and
  lower $U$ may be obtained for even higher metallicities. 

 These results are consistent with the conclusions of
 Rodriguez-Pascual et al. (1997) who argue in favour of smaller
 density and higher ionization parameter in NLSy1 galaxies. 
 Note, however, that they did
 not consider super-solar metallicities in their model.

 The next question is whether the parameters of the warm absorber in
 NGC~4051 are consistent with being in pressure equilibrium with the
 BLR clouds. This is a non-trivial question in the case of NGC~4051
 with highly variable ionizing continuum. Presenting non-equilibrium
 photoionization calculations, Nicastro et al. (1999) estimated that
 the warm absorber in NGC~4051 is at a distance of $\sim$$1.3 \pm
 0.3 \times 10^{16}$ cm from the nucleus, which is  
 similar to that of the BLR. Therefore,
 pressure equilibrium with the BLR clouds is possible. 
A more detailed comparison will be possible, once X-ray observations
of higher spectral resolution of NGC\,4051 become available,
in combination with photoionization calculations that include
time-dependent effects.

\section {Discussion}

\subsection{Evidence for supersolar metallicity in NLSy1 galaxies} 

Mathur (2000a; see also Kawaguchi \& Aoki 2000) suggested 
NLSy1s to be young/re\-ju\-venated active galaxies.
In analogy to high-z quasars they may have super-solar metal
abundances, as indicated by a number of observations:

An overabundance of iron was suggested to explain the strong optical
FeII emission in NLSy1s (Collin \& Joly 2000).  Supersolar Fe or Ne
provide a possible explanation for the peculiar absorption features
around $\sim$ 1 keV seen in some NLSy1 galaxies (e.g., Ulrich et
al. 1999, Turner et al. 1999). Overabundant Fe was also discussed to
explain the strength of the FeK$\alpha$ lines (Fabian 1999).  Nitrogen
may be overabundant as well (e.g., Wills et al. 2000, Mathur 2000b).

\subsection{Properties of the BLR in NLSy1 galaxies} 

X-ray spectra steeper than \G $\approx$ --2.5 completely
remove the multi-phase equilibrium (see Fig. 7 of KM2000), whereas
a number of NLSy1s with even steeper soft X-ray spectra have been observed.
So, within the context of the multi-phase model it is important to
ask, which mechanism could alter that trend. 

In addition, the study of Rodriguez-Pascual et al. (1997) shows that
NLSy1s do have {\em broad UV} lines (FWHM $\simeq 5000$ km/s); only
the optical BLR lines appear smaller (FWHM $\approxlt 2000$ km/s). The
existence of the broad UV lines has to be explained.

We find that  supersolar metal abundances
 halt complete removal of multiple phases in pressure balance,
and therefore allow for the presence of a BLR.

\subsection{Multi-phase BLR model}

Before we discuss consequences of the models considered here,
we first give some cautious comments on the general
modeling assumptions. 
(i)  
It was assumed that there is no strong temperature gradient across the
clouds. For very large column densities that would no longer be true.
(ii) The detailed shape of the equilibrium curve in the
intermediate-temperature region, where metals are important, depends
on the details of the heating-cooling processes and reflects to some
extent the completeness with which these are implemented in the code
used (see Ferland et al. 1998, Kingdon \& Ferland 1999).  (iii) The
EUV spectrum also has an important influence on the shape of the
curves but it has already been studied previously (Fabian et al. 1986,
KM2000).

Furthermore, it has to be kept in mind that a number of alternative
BLR models have been studied.  The BLR clouds might be confined by
mechanisms other than thermal pressure (e.g. magnetic fields could
play a role; Rees 1987), or they might not be confined at
all. Confinement is unnecessary if the BLR clouds are continuously
produced{\footnote{e.g., Murray \& Chiang 1995 (disk-driven winds; see
also Dietrich et al. 1999), Edwards 1980, Alexander \& Netzer 1994
(bloated stars), Perry \& Dyson 1985 (cloudlet creation by radiative
shocks)}}.  These alternatives to cloud-confinement models face
problems as well, though.  For instance, BLR cloud counting arguments
made the bloated stars and related scenarios unlikely (at least as
explanation for the low-ionization component of the BLR; Dietrich et al. 1999).

With all this in mind, we come back to variants of the
Krolik-McKee-Tarter (KMT) model. 
It is interesting to note that recent {\sl Chandra} observations 
gave further input to the `classical' confinement scenarios,
based on the detection of extended high-temperature X-ray
emission in an AGN (Ogle et al. 2000), interpreted as hot intercloud
medium.   

 The study of Rodriguez-Pascual et al. (1997) showed 
that NLSy1 galaxies possess {\em broad} emission lines
in the UV. They explained the different line-widths
measured in the optical and UV by a {\em matter-bounded} BLR,
as discussed by Shields et al. (1995), 
in NLSy1 galaxies. 
This fits to the KMT model approach, which predicts a 
thin shell-like BLR.

\subsection{Warm absorbers in NLSy1 galaxies}

The region in which the temperature is multi-valued for constant
pressure is preserved by supersolar metallicity in NLSy1s. If the X-ray
spectra are not too steep, a pressure balance between
a cold phase (the BLR in NLSy1s) and a hotter intermediate phase is
possible (even if the very hot phase is absent).

This intermediate region has previously been associated with
ionized absorbers (e.g., Reynolds \& Fabian 1995, Komossa \& Fink 1997a,b,
Elvis 2000a,b; see Fig.\,1 of Komossa 2001 for a summary). 

The suggestion that NLSy1s are young objects with a high accretion
rate makes the presence of a large amount of (ionized) gas in the
nucleus likely (Mathur 2000b).  Indeed warm absorbers have been
detected in quite a number of NLSy1 galaxies (e.g., 
Brandt et al. 1997, Komossa \& Fink 1997a, Hayashida 1997,
Komossa \& Bade 1998, Iwasawa et al. 1998, Ulrich et al. 1999, Vaughan et
al. 1999), or have been suggested to explain some unusual
spectral/variability properties of NLSy1s (e.g., Komossa \&
Meerschweinchen 2000).  Some studies indicate that ionized absorbers
are somewhat less abundant in NLSy1s than in Seyferts (Leighly 1999).
The reason could be a {\em higher} degree of ionization of the
absorbers 
which makes them less easily detectable in NLSy1s.  The
OVII absorbers, seen in some NLSy1 galaxies (e.g., IRAS 17020+4544),
which might at first glance suggest a low degree of ionization, 
can
easily be explained by the presence of a {\em second} 
warm absorber component which is located at larger distances
from the nucleus, and which is mixed with dust
(Komossa \& Bade 1998).

\section {Summary and conclusions}

We have presented a study of the influence of metallicity on the
multi-phase equilibrium in photoionized gas. We find that supersolar
metallicity increases the range over which a multi-phase medium in
pressure balance may exist. In objects with steep X-ray spectra, like
NLSy1s, such an equilibrium is not possible if gas metallicity is less
than or equal to solar. As a result a pressure confined BLR cannot
exist. This problem is alleviated if metallicity is supersolar. This
offers further credence to the suggestion that NLSy1s may have
supersolar metallicity.  High resolution X-ray spectroscopy, together
with optical/UV spectroscopy, is required to derive elemental 
abundances and to
address these issues in more detail.
The results of this {\sl Note} are also of relevance to high-redshift quasars,
since their BLR clouds show supersolar metal abundances.

\begin{acknowledgements}
This work is supported in part by the NASA grant NAG5-8913 (LTSA) to SM.
We thank 
Gary Ferland for providing {\em Cloudy}.   
Preprints of this and related papers can be retrieved  
at http://www.xray.mpe.mpg.de/$\sim$skomossa/
\end{acknowledgements}

\end{document}